\documentclass{jaa}
\usepackage[english]{babel}
\usepackage{natbib, amsmath, rotating}

\def\amp
\def\pasa{PASA}

\def\lsim{~\rlap{$<$}{\lower 1.0ex\hbox{$\sim$}}}

\def\gsim{~\rlap{$>$}{\lower 1.0ex\hbox{$\sim$}}}

\newcommand{\HI}{{H{\sc i}}}

\usepackage{graphics}
\usepackage{graphicx}
\usepackage{color}
\usepackage{amsmath}
\usepackage{psfrag}
\usepackage{epsfig}
\usepackage{subfigure}

\begin{document}
\title[Prowess]
{Prowess - a software model for the Ooty Wide Field Array}
\author[Marthi] {Visweshwar Ram
  Marthi\thanks{Email:vrmarthi@ncra.tifr.res.in}\\
     National Centre for Radio Astrophysics, Tata Institute of
  Fundamental Research, \\Post Bag 3, Ganeshkhind, Pune - 411 007, India. \\
    }
\date {}
\maketitle

\begin{abstract}
One of the scientific objectives of the Ooty Wide Field Array(OWFA) is to
observe the redshifted \HI~emission from $z \sim 3.35$. Although predictions
spell out optimistic outcomes in reasonable integration times, these studies
were based purely on analytical assumptions, without accounting for limiting
systematics. A software model for OWFA has been developed with a view to
understanding the instrument-induced systematics, by describing a complete
software model for the instrument. This model has been implemented through a
suite of programs, together called \textbf{Prowess}, which has been conceived
with a dual role of an emulator as well as observatory data analysis
software. The programming philosophy followed in building \textbf{Prowess}
enables any user to define an own set of functions and add new
functionality. This paper describes a co-ordinate system suitable for OWFA in which
the baselines are defined. The foregrounds are simulated from their angular
power spectra. The visibilities are then computed from the foregrounds. These
visibilities are then used for further processing, such as calibration and power
spectrum estimation. The package allows for rich visualisation features in multiple
output formats in an interactive fashion, giving the user an intuitive feel for
the data. \textbf{Prowess} has been extensively used for numerical predictions
of the foregrounds for the OWFA \HI~experiment.
\end{abstract}

\newpage
\section{Introduction}
One of the principal aims of the upgrade of the Ooty Radio Telescope (ORT) to OWFA\citep{Prasad2011,
  Subrahmanya2017a, Subrahmanya2017b} is to enable the detection of \HI~emission from large scale 
structures in the post-EoR universe at redshifts $\sim 3$. Theoretical 
calculations of the expected emission tuned to the projected parameters of 
OWFA indicate that the telescope should have sufficient sensitivity
to detect the power spectrum of the redshifted  \HI~21-cm emission, in integration times
of a few hundred hours\citep{Ali2014, Bharadwaj2015, Sarkar2017a}. 
Currently, a number of
experiments are in various stages of progress that aim to directly
detect the brightness temperature
fluctuations $\delta T_b$ of the 21-cm post-reionisation cosmological signal
like the Canadian Hydrogen Intensity Mapping Experiment (CHIME;
\citealt{Bandura2014}), Baryon Acoustic Oscillation Broadband and Broad-beam
Array (BAOBAB; \citealt{Pober2013b}) and
the Tianlai Cylinder Radio Telescope (CRT; \citealt{Chen2011, Xu2015}).
These experiments would each operate at different frequency ranges; BAOBAB 
has been proposed to specifically detect the Baryon Acoustic Oscillation (BAO) feature in the redshifted \HI~21-cm line in the
600-900 MHz band. The Tianlai CRT also is gearing up to
detect the BAO features and constrain dark energy through redshifted \HI~21-cm
observations in the 700-1400 MHz band\citep{Chen2011, Xu2015}. CHIME would
overlap with both these experiments in the range $\sim 400-800$ MHz. The OWFA
cosmology experiment is expected to fill a significant gap in understanding the
evolution of post-reionisation neutral hydrogen at large scales in an important redshift
interval of $z\sim 3.35$. 

While the raw sensitivity of the telescope would be sufficient to detect
the \HI~emission from redshifts $z\sim 3.35$\citep{Ali2014, Bharadwaj2015}, the expected signal
is many orders of magnitude fainter than the other astrophysical signals,
i.e. the ``foregrounds''\citep[see e.g.][]{Santos2005, Ali2008}. These include 
emission from the diffuse ionised galactic interstellar medium (``diffuse Galactic
synchrotron emission'' and ``galactic free-free emission'') and emission from 
the extragalactic radio sources(called ``the
extragalactic foreground'') that the telescope is sensitive to.
Many instrumental effects come into play when the signal of interest is
buried several in orders of magnitude brighter foregrounds:
systematics introduced by uncalibrated antenna gains, interference
from terrestrial sources and effects of the complex intrinsic interaction between the
instrument and the foregrounds, due to the chromatic response of the telescope, are
thought to be the dominant contributors.
A good understanding of all of these issues is required to enable a robust
prediction of the cosmological signal that could be detected through
observations with OWFA.

The first step to understanding the instrumental systematics is to develop a thorough understanding of
the instrument itself, and capture it in a software model
that would include all the expected instrumental effects. All of the
astrophysical signals(e.g. the diffuse Galactic and the extragalactic point
source foregrounds) can then be suitably parametrized and 
included in the model. One of the expected by-products of the exercise to detect the 
cosmological \HI~signal is a better understanding
of some of the foregrounds, particularly of the diffuse Galactic foreground, which is the dominant foreground from within the Galaxy
and of interest in itself\citep[see e.g.][]{Iacobelli2013a, Iacobelli2013b, Iacobelli2014}. 
The ability to characterise the foregrounds and the fundamental limitations set by the
instrument are both crucial to enable realistic predictions for the redshifted
\HI~21-cm detection. The software model described in this paper was developed
in order to help better understand
the systematics, as well as devise methods to devise foreground characterisation
and subtraction methods.

\section{The rationale for a software model}\label{sec:rationale}
The OWFA \HI~experiment is a
challenging one in terms of both the special hardware requirements as
well as the methods and algorithms that eventually enable us to
measure the \HI~power spectrum. A  significant component of the design
of an experiment, especially in modern low frequency radio cosmology,
has been the investment in simulating the
instrument and the experiment itself based purely on a software
model. The results from simulations can often
influence the course of the experiment through valuable insight. This
has been the driving philosophy for an emulator based on a software
model for the OWFA \HI~experiment. 

For OWFA, traditional interferometric data analysis software packages are not useful as they do not provide
sufficient functionality for redundancy
calibration, or for the final processing which, in this case, is not imaging.
A complete software suite has been developed consisting of several
standalone programs that serve two simultaneous purposes:
\begin{itemize}
\item to simulate visibilities as obtained from OWFA, based on the
  instrument and sky description. The instrument description should
  naturally lead to all the effects and systematics that are 
  expected to be present in an actual interferometric
  observation. Simulated data can provide a test bed for devising
  and refining RFI mitigation algorithms, calibration algorithms and
  statistical estimators for signal characterisation, etc.
\item to function as the observatory software pipeline that is used to
  process real data from the telescope, so that the
  simulations above inform us to refine and adopt the optimal
  strategies for working with real data.
\end{itemize}

Given that a software model for the telescope and
the sky would serve as a very useful guide to 
the experiment, I shall now set up the preliminaries for describing the
emulator in some detail, and list its various features with examples.

\section{The programming philosophy}
The software suite that has been developed for OWFA
simulations is a C-based collection of utilities and algorithm
implementations. Visualisation of data is almost
the first step in data handling and a standard format definition
should therefore be the first choice.
%Infact, it serves the dual purpose of an
%emulator and the observatory package for handling post-correlation
%data. 
%-------------------------
%It would be shortly clear that it is not
%unreasonable to call the package an emulator, given the fact that it
%has greatly influenced and shaped our understanding of the instrument
%and the systematics we can expect at OWFA in the cosmology
%experiment. 
%----------------------------
%Infact, our study of these systematics forms a substantial
%portion of this thesis, and these would appear in later chapters.
The suite was conceived from the early days as one that would grow
organically to accommodate observatory needs. Therefore, the choice
was to adhere to an international standard for the visibility data, so
that a team of astronomers stationed in widely separated geographical
locations can handle these data using this software suite.
The following considerations were kept in mind during the development of the emulator.
\begin{itemize}
\item The Flexible Image Transport
System(FITS; \citealt{FITS}) or the Measurement Set(MS; \citealt{MS})
format definitions were the obvious formats to choose
from. %A FITS file can be converted to MS or HDF5 using readily available tools. 
%Given
%the popularity that FITS enjoys and the number of FITS tools
%available, like cfitsio, ds9 and fv(to name a few), it was but natural
%to choose it. Therefore the programs were developed around the FITS
%format for easy data portability. Besides, 
The fact that FITS is the data format
at the GMRT and hence is familiar to astronomers both within NCRA and users of
the GMRT played a significant role in the decision to adopt it. However, it is
worth mentioning that the Hierarchical Data Format (HDF) \citep[see e.g.][for
  advantages of HDF5 over UVFITS]{Price2015} could be considered in the future due to its advantages over FITS, such as
compressibility and parallel I/O. The availability of conversion libraries from
FITS to HDF5 makes this an attractive alternative if the full potential of
parallel processing is to be harnessed. Pre-conversion of UVFITS observatory
data to HDF5 is then the only extra step in the offline processing necessitated
by this choice. UVFITS will therefore currently remain the
observatory data format of choice. 

\item The second major factor was the ability to add new
utilities to the suite by anyone familiar with the FITS
definition. Therefore, the suite comes with a very rich set of
subroutine functions that do standard operations in a transparent
manner. A new utility that has to operate on the FITS data at the
level of individual records can hence draw on this library of
subroutines. 

\item Finally, a high level of ease with which utilities can be added
to the suite is achieved by making the chore of passing arguments to
function calls a trivial operation. Instead of passing specific
arguments to functions, the pointer to a global superstructure is
passed uniformly to all function calls. The superstructure itself is a
structure of many structures, which are defined in different header
files depending on their functionality. Therefore, any user would be
able to add her own function definition to the existing
subroutines by passing a single argument.
\end{itemize}

\begin{figure}
\begin{center}
\includegraphics[scale=0.33]{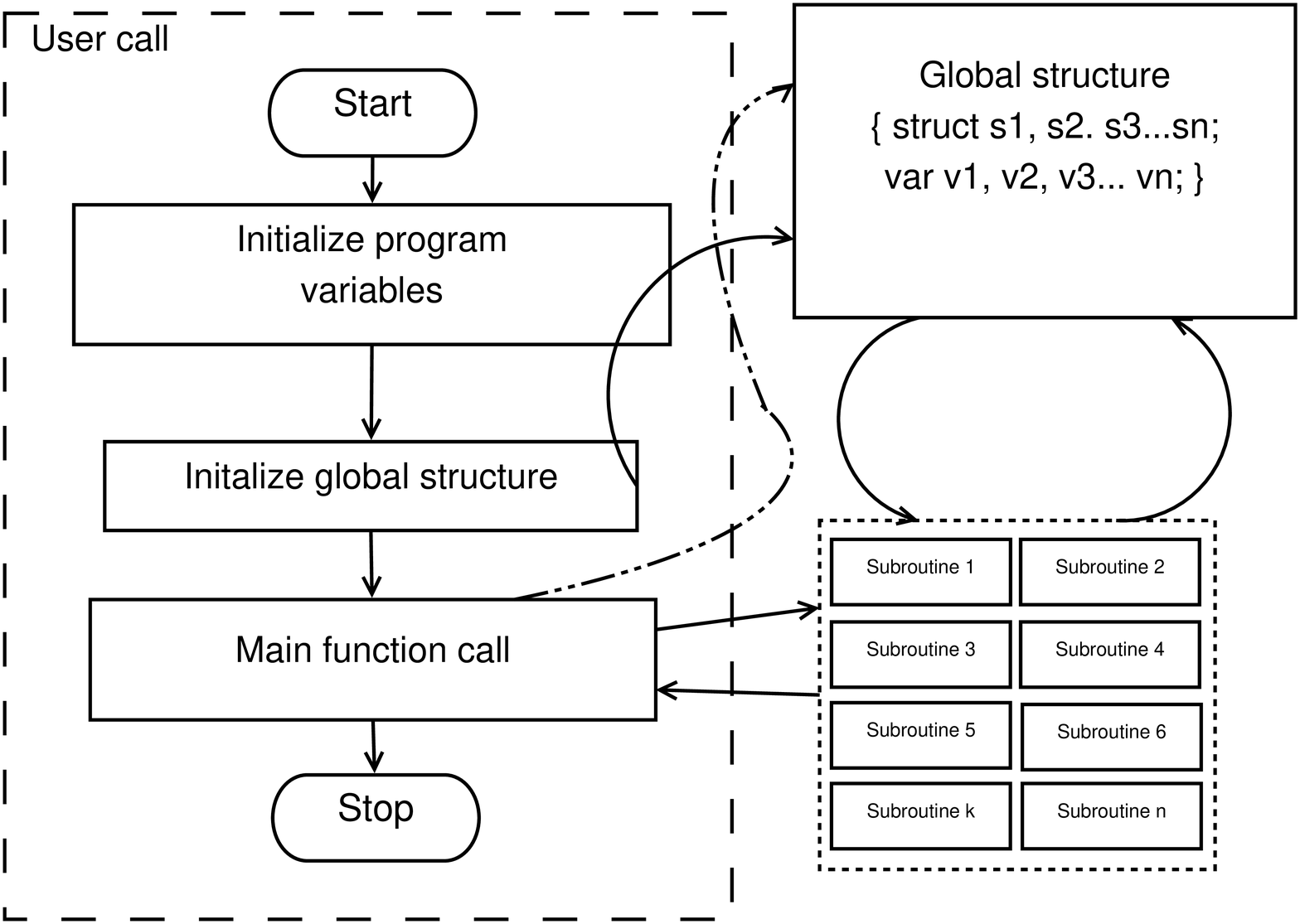}
\end{center}
\caption{The general model for each program in the OWFA simulator
  suite}
\label{fig:philosophy}
\end{figure}

Figure~\ref{fig:philosophy} captures the spirit of the programming
philosophy. The main call merely initialises a few variables
specific to the program. 
Subroutines, variables and structures are segregated according to
their functionality and defined in appropriate header
files. Similarly, subroutine functions are defined in functionally
separable C program files. For example, all function definitions
related to the instrument are available in the \textbf{inc/sysdefs.h} and
\textbf{src/syssubs.c} files. FITS-related definitions and structures
are to be found in \textbf{inc/fitsdefs.h} and \textbf{src/fitssubs.c}
files. Sky simulations are grouped under \textbf{inc/skydefs.h} and
\textbf{src/skysubs.c}. Mathematical function definitions and
structures are grouped under \textbf{inc/matdefs.h} and
\textbf{src/matsubs.c}. The global superstructure, which holds all the variables
and structures is defined to be of type \textbf{ProjParType}, which is defined
in \textbf{inc/sysdefs.h}\footnote{\textbf{Prowess} is continually evolving, but
  is available on request.}.

\section[Prowess - OWFA emulator]{\textbf{Prowess} - a \textbf{Pr}ogrammable \textbf{OW}FA \textbf{E}mulator \textbf{S}y\textbf{s}tem}
The software suite is given the name ``Programmable OWFA Emulator System'', and it is self-explanatory. Programmable, because
adding new utilities or functionality is made easy as described in the
previous section, and the emulator is specific to OWFA. I shall now describe the preliminaries required to capture the
instrument in a software model.
\subsection{Antennas and baselines}
OWFA\citep{Subrahmanya2017a, Subrahmanya2017b} would operate
in two concurrent modes - Phase-I and Phase-II. Phase-I is a 40-antenna interferometer and Phase-II is a
264-antenna interferometer. Only Phase-II refers to an operational system and
Phase-I is achieved in software. The Ooty Radio Telescope(ORT)
is a $\sim 530$ m long cylinder that is 30 m wide\citep{Swarup1971}.
The telescope consists of 1056 dipoles arranged regularly along its
length. These 1056 dipoles are grouped into 22 modules, each module being
supported mechanically by a parabolic frame. All the 22 frames are steered in
unison through a common drive shaft. Each of the 22 modules is the sum of a
contiguous group of 48 dipoles, combined through a passive combiner network. The
signals from the dipoles are summed hierarchically, and the second smallest unit
in this hierarchy is the output of the 4-way combiner\citep{Subrahmanya2017b}. 
The signals from six Phase-II apertures are again summed in two stages to give the Phase-I aperture.
Correspondingly, the two interferometer modes provide two different aperture settings:
\begin{itemize}
\item The output of the 4-way combiner forms the aperture of the Phase-II
  system. Every group of 4 dipoles, or a sixth of a half-module, would operate as a
  single element in Phase-II. This corresponds to 1.92 m of the 530 m
  long cylinder, equivalent to 2$\lambda$. This results in 264
  apertures throughout the length of the telescope.

\item The output of the sum of six 4-way combiners forms the aperture of the Phase-I
  system,  equivalent to one half-module. Therefore, every group of 24 dipoles
  would operate as a single element in Phase-I. This corresponds to 11.5 m of the 530 m
  long cylinder, equivalent to 12.5$\lambda$. This results in 40 apertures
  throughout the length of the telescope.
\end{itemize}

\begin{figure}
\begin{center}
\begin{verbatim}
#AntId	   Ant	    bx	    by      bz     ddly    fdly      
#------------------------------------------------------------
ANT00	   N10N	   0.0    0.0    10.0	     0.0     0.0
ANT01	   N10S	   0.0	   0.0    21.5	     0.0	    0.0
.        .       .      .      .         .       .
.        .       .      .      .         .       .
.        .       .      .      .         .       .
ANT38	   S10N	   0.0	   0.0	  447.0	     0.0	    0.0
ANT39	   S10S	   0.0	   0.0	  458.5      0.0	    0.0
#END
\end{verbatim}
\end{center}
\caption{The antenna definition file Antenna.Def.40 for Phase-I of the
  OWFA interferometer.}
\label{fig:ant-def}
\end{figure}
To begin with, the emulator has to be initialised with the
antenna positions. This is done through an input
Antenna Definition file, ``Antenna.Def.40'' for Phase-I and
``Antenna.Def.264'' for Phase-II. The parsing section of the code then
figures out which of the two modes the telescope is being operated
in. Accordingly, it sets the aperture dimensions. The user has the
option to switch off certain antennas in the ``Antenna.Def'' file to
simulate a situation when some antennas are not available. These are then omitted from the
simulations as well as the output visibility data. This
not only obviates the need to maintain a running log of the invalid
antennas, but also eases memory and storage requirements. 
Figure~\ref{fig:ant-def} shows the antenna definition for Phase-I. The
file has seven columns: 
\begin{enumerate}
\item Column 1 shows the antenna identifier used by the FITS standard.
\item Column 2 is the antenna name; the name of each antenna is tied
  to its identifier. This helps in unambiguous bookkeeping even when
  switching off certain antennas.
\item Columns 3, 4 and 5 respectively give the antenna $x, y$ and $z$
  co-ordinates in a right-handed co-ordinate system, which shall be described shortly.
\item Columns 6 and 7 respectively denote the delay in seconds, corresponding to
  the number of integer and fractional clock cycles offset with respect to a reference antenna. At the moment, these fields are not being used,
  hence their values are all set to zero. In practice, these would
  represent the fixed delays arising from differences in cable lengths.Therefore
  they can be measured reasonably accurately and are unlikely to change in an
  undisturbed setup. These fields would be used to correct for the phase ramps
  in the visibilities across the band for each baseline.
\end{enumerate}

Since the dipoles are regularly spaced, the equivalent apertures in Phase-I and
Phase-II are also regularly spaced. This results in an interferometer in which
the separation between any pair of apertures is an integral multiple of the
shortest separation between adjacent apertures, 
\begin{equation}
d_{\mathrm{n}} = nd
\label{eqn:physical-baseline}
\end{equation}
where $d$ is the both the size of the aperture as well as the shortest
spacing. The 40-antenna Phase-I has twenty half-modules in the northern half and twenty in
the southern half. The northern ones are
named N01 to N10 outwards from the midpoint of the telescope, and
similarly the southern modules. The two half-modules within each
module are given a ``N'' or ``S'' identifier. The antenna definition
file is parsed and the values are stored in the antenna structure
within the superstructure(\textbf{ProjParType}). 

In the co-ordinate system chosen for OWFA,
the antennas are placed along the $z$-axis. 
%Since the absolute
%positions of the antennas make no difference to the baselines in an
%interferometer, we can arbitrarily set any antenna as the reference.
\begin{figure}
\begin{center}
\begin{verbatim}
###  Antenna  	Re(gain)   	Im(gain)    abs(gain)    arg(gain)
###
 0   	N10N   	 2.111527   -1.651020   	2.680375   	-38.022146
 1   	N10S   	-0.769040    2.448255    2.566198   	107.438412
 .    .        .           .           .            . 
 .    .        .           .           .            . 
 .    .        .           .           .            . 
38   	S10N   	 0.112905    1.564518    1.568587   	 85.872353
39   	S10S   	 1.386609   -1.809840   	2.279958   	-52.542475
\end{verbatim}
\end{center}
\caption{The antenna initialiser log, with the complex gain assigned
  to each antenna.}
\label{fig:ant-init.info}
\end{figure}
Each antenna $i$ is assigned a complex, frequency dependent, electronic gain $g_i$, obtained
as a random complex number distributed around a mean gain $|g|$,
referred to the central frequency $\nu_0$ = 326.5 MHz. The observing
band is split into $N$ channels. It is possible to introduce slowly time-varying
gains in the simulations about the mean antenna gains described above, but such
gain variations are automatically handled at the time of calibration.
\begin{figure}
\begin{center}
\begin{verbatim}
###	   Ant1  Ant2  	 FITSbl  Ant1    Ant2
###
  1   	 0   	 1   	   257   	N10N   	N10S
  2   	 0   	 2   	   258   	N10N   	N09N
  .     .     .       .      .       .
  .     .     .       .      .       .
  .     .     .        .     .       .
 38   	 0   	38   	   294   	N10N   	S10N
 39   	 0   	39   	   295   	N10N   	S10S
\end{verbatim}
\end{center}
\caption{The log ``baselines.info'' that lists the baselines counted
  as pairs of the available antennas. Only the first 39 baselines of
  Phase-I are shown as an example.}
\label{fig:baselines.info}
\end{figure}
The antenna structure initialization information is written out
in a log called ``ant-init.info''. The real and imaginary parts of the
complex gains, as well as its amplitude and phase(in degrees) at $\nu_0$
are written out in the file. An example is shown in Figure~\ref{fig:ant-init.info}.

Baselines are then obtained as antenna pairs, and each baseline is
written into a structure that holds the baseline number, the
participating antenna pair, the three dimensions of the baseline and its length
in wavelength units at the reference frequency. A log of the baselines is
written out to ``baseline.info'', part of which is shown in
Figure~\ref{fig:baselines.info}. An $N_A$ antenna interferometer results in $^{N_A}C_2$
baselines, giving 780 for Phase-I and 34716 for Phase-II.
The baseline vectors are obtained from the physical antenna
separations, defined at the central frequency $\nu_0$
but at each channel it is appropriately scaled when computing the
visibilities.
\begin{equation}
\mathbf{d}_{|a-b|} = \mathbf{x}_a - \mathbf{x}_b
\label{eqn:baseline-phy}
\end{equation}
\begin{equation}
\mathbf{U}_{|a-b|} = \mathbf{d}_{|a-b|}\ \frac{\nu}{c}
\label{eqn:lambda-baseline}
\end{equation}
Equation~\ref{eqn:baseline-phy} shows the physical separation between
antenna pairs, whereas equation~\ref{eqn:lambda-baseline} shows the baseline in
wavelength units at any given frequency $\nu$. 
The regular spacing of the antennas results in baselines with
redundant spacings. As a result, we obtain $N_A-n$ copies of the baseline with a
separation of $n$ units.
In this case of an $N_A$-antenna linear array, only
$N_A-1$ baselines out of $^{N_A}C_2$ are unique and non-redundant. All of these
$N_A-1$ baselines have redundant copies, except the longest one.
\subsection{A co-ordinate system suitable for OWFA}
\begin{figure}
\begin{center}
\includegraphics[scale=0.6]{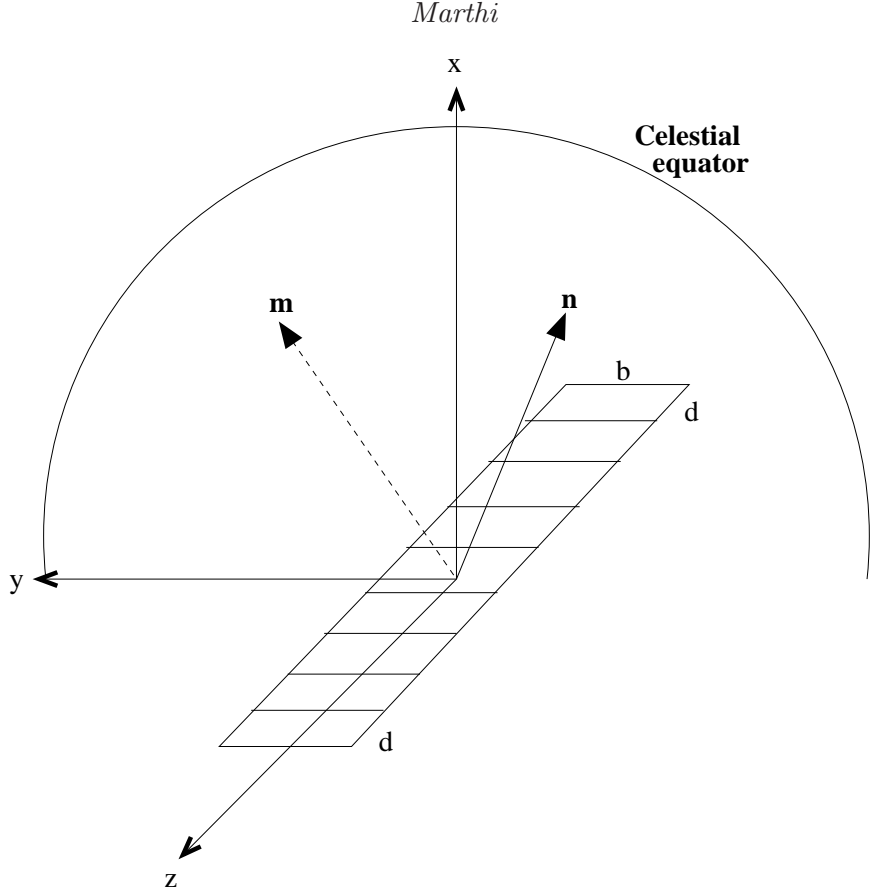}
\caption{A schematic of the co-ordinate system for computing the visibilities,
  in which $\mathbf{n}$ is an arbitrary direction and $\mathbf{m}$ is the
  direction of pointing. The visibilities are computed over the entire solid
  angle of the celestial hemisphere.}
\label{fig:owfa_coord}
\end{center}
\end{figure}

A generalised framework for computing the visibilities is presented here.
Consider a right-handed Cartesian coordinate system, shown in Figure~\ref{fig:owfa_coord},
tied to the telescope, in which the $z$-axis is along the N-S direction, parallel to the
axis of the parabolic cylinder, the $x$-axis is aligned with the normal to the
telescope aperture which is directed towards $(\alpha_0,0)$ on the celestial
equator, and the $y$ axis is in the plane of the telescope's aperture, perpendicular
to both the $x$ and $z$ axes respectively. $\mathbf{\hat{i}}, \mathbf{\hat{j}}$ and $\mathbf{\hat{k}}$ denote the
unit vectors along $x, y$ and $z$. In this coordinate system, we have     
\begin{equation}
\mathbf{U}= v \mathbf{\hat{k}}
\end{equation}
Observations are centered on a position 
$(\alpha_0,\delta_0)$  on the celestial sphere: let the unit vector $\mathbf{\hat{m}}$
denote this position on the celestial sphere. $\mathbf{\hat{m}}$ always lives on the
$x\!-\!z$ plane, and is given by
\begin{equation}
\mathbf{\hat{m}}=\sin(\delta_o)  \, \mathbf{\hat{k}} + \cos(\delta_0)   \, \mathbf{\hat{i}}
\end{equation}
 The measured visibility for a baseline of length $\mathbf{U}$ at a frequency $\nu$
 can be written as 
\begin{equation}
\mathbf{M}(\mathbf{U},\nu)=\int d \Omega_{\mathbf{\hat{n}}} \, I(\mathbf{n},\nu)  \,  A(\Delta \mathbf{n},\nu) \, 
 e^{2 \pi i \mathbf{U} \cdot \Delta \mathbf{n}} 
\label{eqn:model_RIME}
\end{equation}
where $\mathbf{\hat{n}}$ refers to an arbitrary direction in the celestial sphere,
given by
\begin{equation}
\mathbf{\hat{n}}=\sin(\delta)  \, \mathbf{\hat{k}} + \cos(\delta) \left[ \cos(\alpha-\alpha_0) \, \mathbf{\hat{i}} + 
\sin(\alpha-\alpha_0) \mathbf{\hat{j}} \right]
\end{equation}
The solid angle integral here is over the entire celestial hemisphere, and 
\begin{equation}
\Delta n = \mathbf{\hat{n}} - \mathbf{\hat{m}} 
\end{equation}
We finally have 
\begin{equation}
\mathbf{M}(\mathbf{U},\nu)=\int [d \sin(\delta) \,  d \alpha ] \,
I(\alpha,\delta, \nu) \, e^{2 \pi i v  [\sin(\delta-\delta_0)] }  A(\Delta n_y,\Delta n_z)
\end{equation}
where 
$\Delta n_y=\cos(\delta) \, \sin(\alpha-\alpha_0)$ and 
$\Delta n_z= \sin(\delta-\delta_0)$. 
Note that in this co-ordinate system the argument of the exponent depends only
on the baseline length and the declination, reflecting the 1D geometry of OWFA.
\subsection{Aperture and the primary beam}
\begin{figure}
\begin{center}
\begin{minipage}{190mm}
\subfigure[$\delta_0 =
  0^\circ$]{\includegraphics[scale=0.25,angle=270]{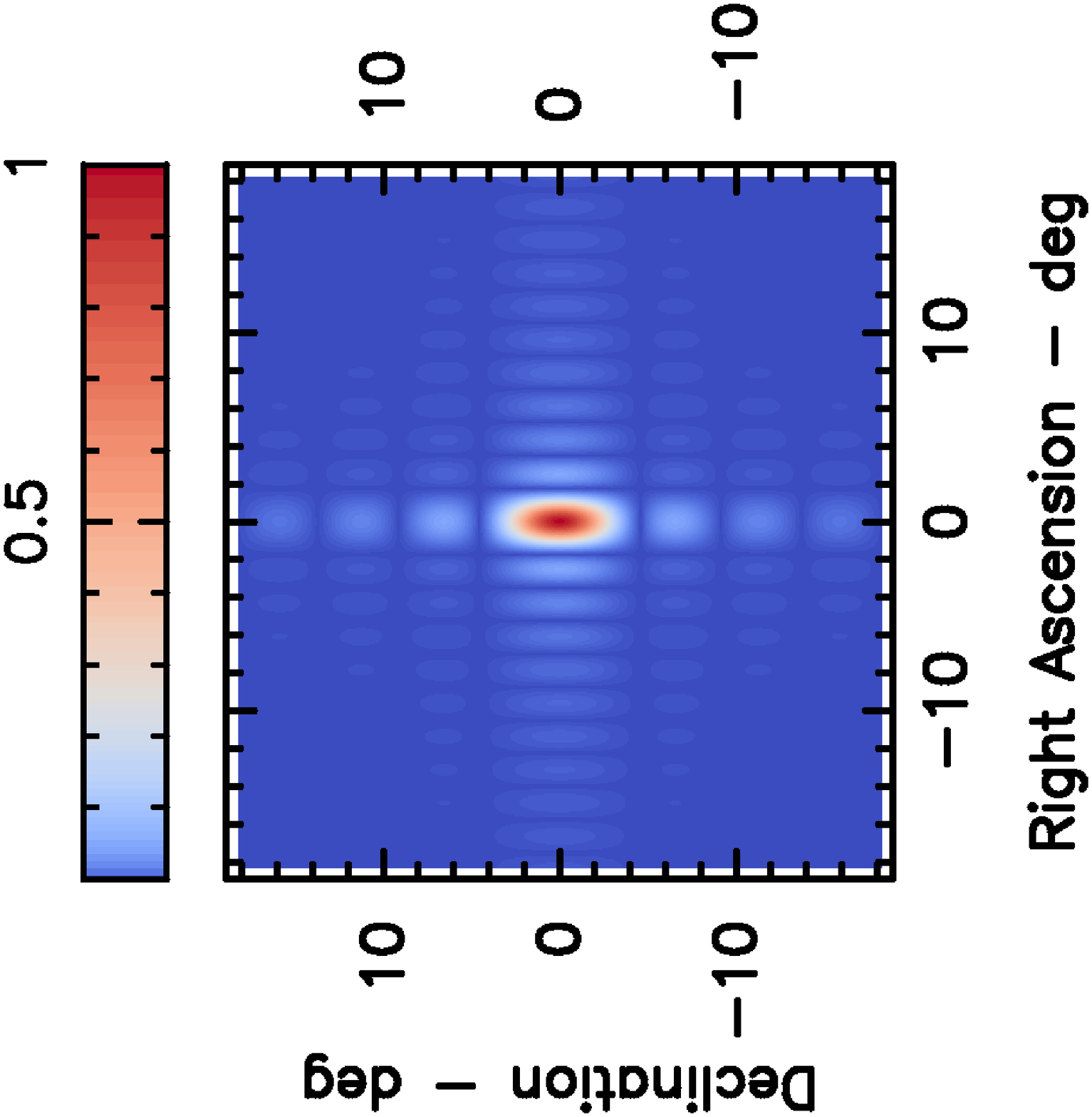}}
\hskip 0.3in
\subfigure[$\delta_0 =
  20^\circ$]{\includegraphics[scale=0.25,angle=270]{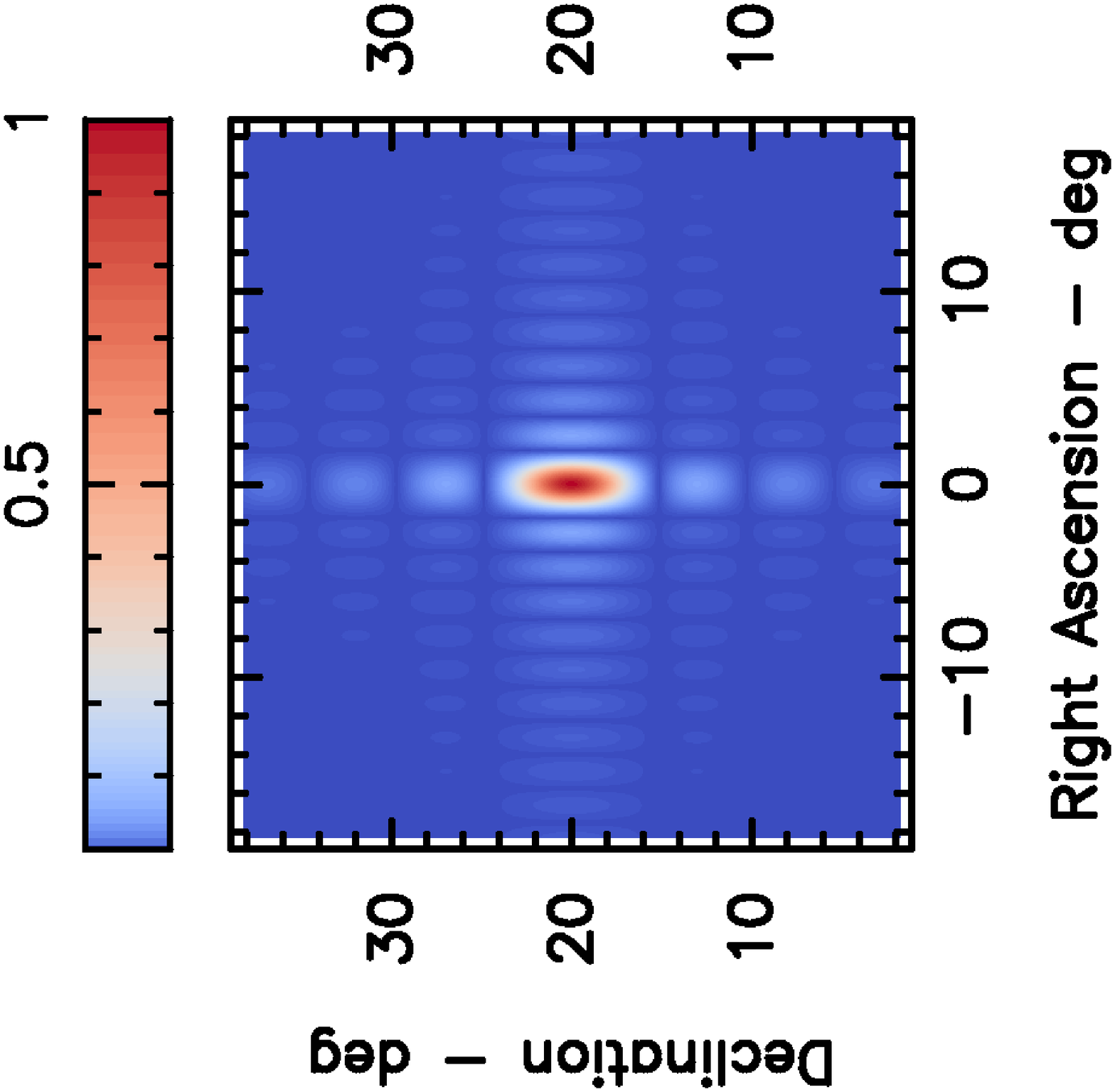}}\\
\vskip 0.3in
\subfigure[$\delta_0 =
  40^\circ$]{\includegraphics[scale=0.25,angle=270]{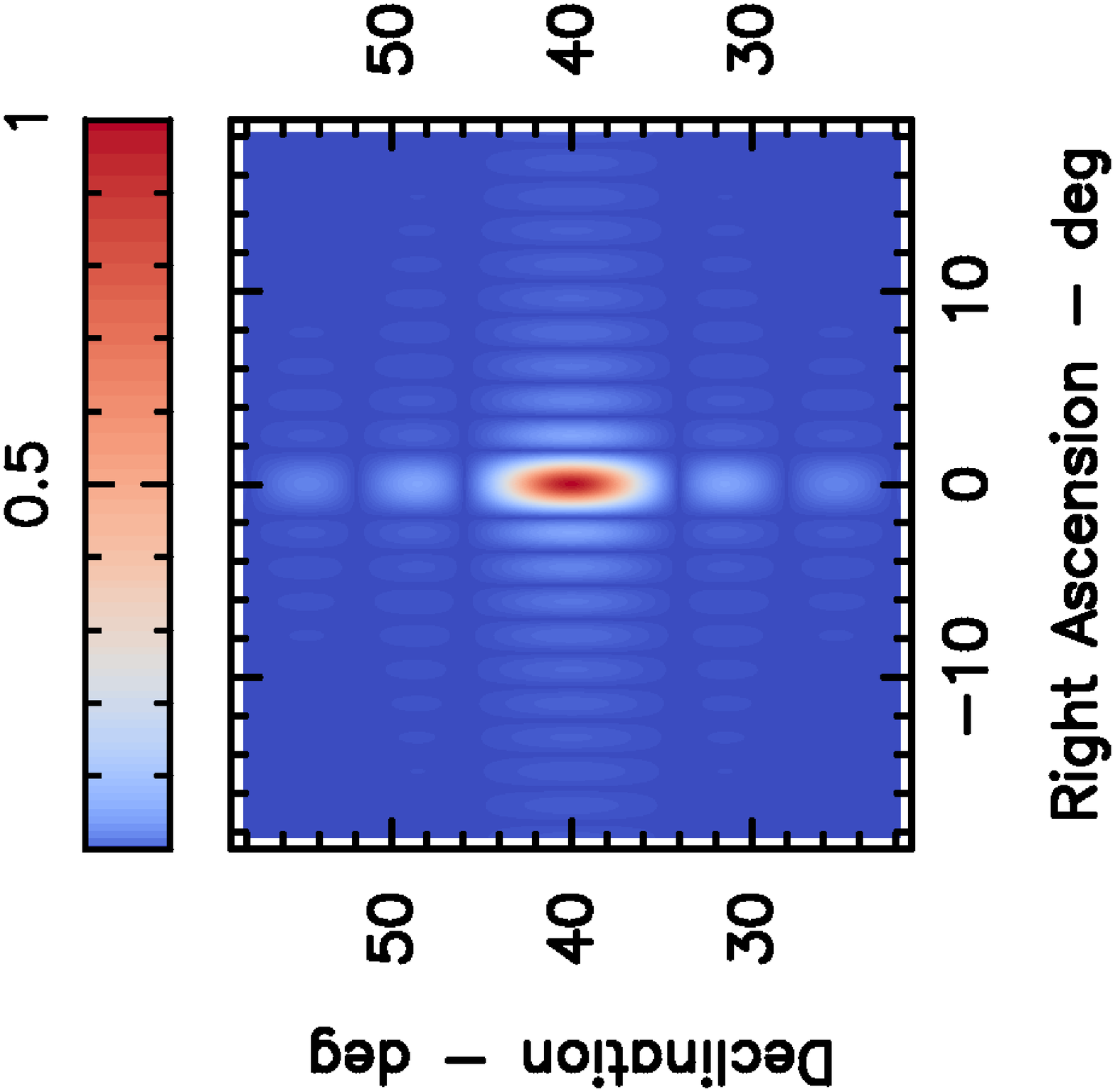}}
\hskip 0.4in
\subfigure[$\delta_0 =
  60^\circ$]{\includegraphics[scale=0.25,angle=270]{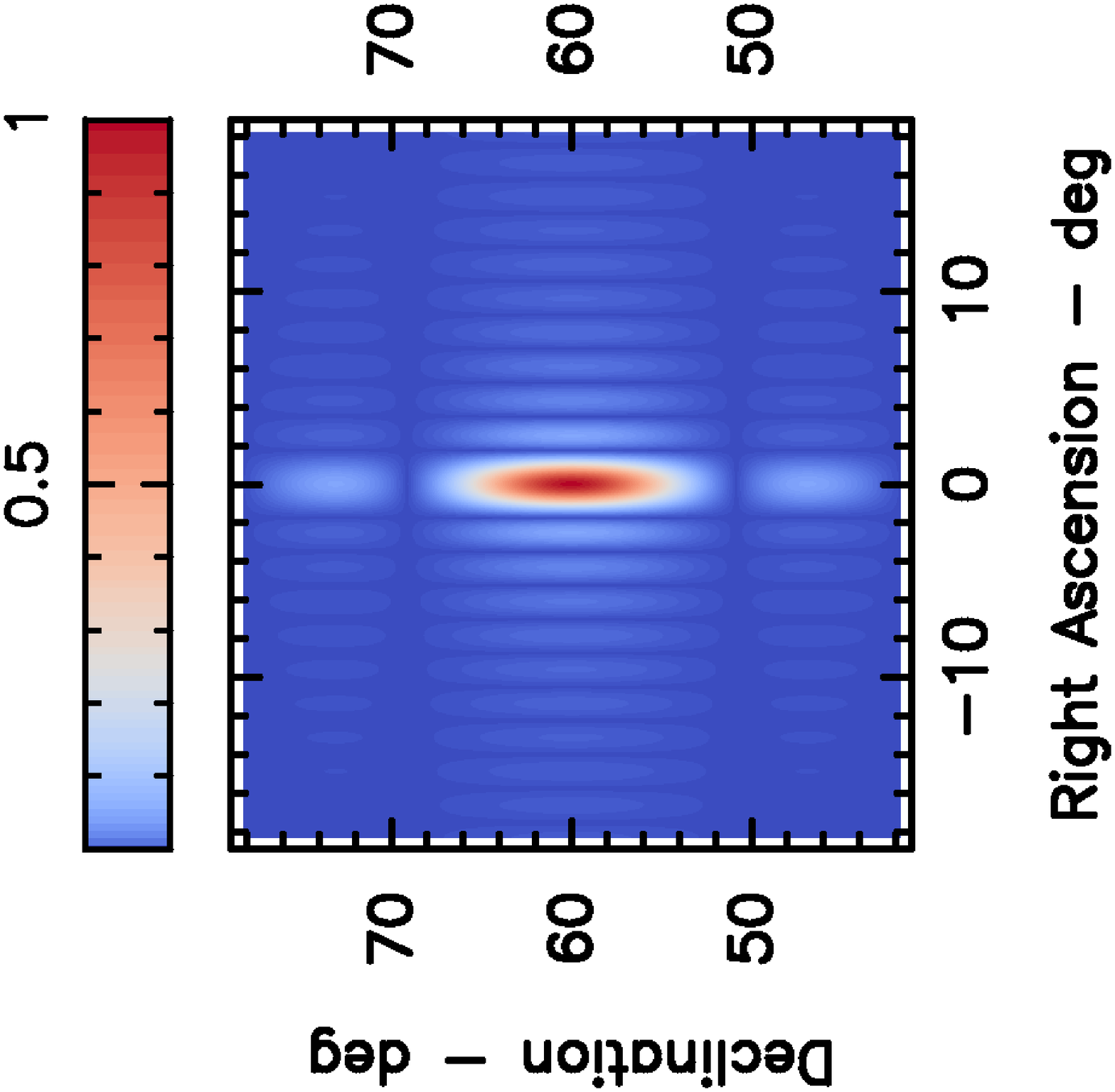}}
\end{minipage}
\end{center}
\caption{The primary beam power pattern at declinations (a) $\delta_0
  = 0^\circ$, (b) $20^\circ$, (c) $40^\circ$ and (d) $60^\circ$. The
  beam widens noticeably in declination extent at higher declination
  as the projected aperture size shrinks as $\cos\ \!\delta_0$.}
\label{fig:PRBEAM-vs-dec}
\end{figure}

We may write the general beam pattern for a rectangular aperture as $A(\Delta \mathbf{n})\equiv 
A(\Delta n_y,\Delta  n_z)$ where $(\Delta n_y,\Delta  n_z)$  are respectively
the $y$ and $z$ components of $\Delta \mathbf{n}$.  The Phase-I
aperture is $11.5 \mathrm{m} \times 30 \mathrm{m}$ and the Phase-II
aperture is $1.92 \mathrm{m} \times 30 \mathrm{m}$ in $d \times b$.
For the rectangular aperture approximated in this paper for OWFA, if we assume for the
moment uniform illumination, $A(\Delta n_y,\Delta n_z)$ can be modelled as a
product of ${\rm sinc}^2$ functions:
\begin{equation}
A(\Delta n_y,\Delta n_z)={\rm sinc}^2 \left( \frac{\pi b\, \Delta n_y}{\lambda} \right)  {\rm sinc}^2 \left( \frac{\pi d \, \Delta n_z }{\lambda} \right)
\end{equation}
However, the primary beam should infact be written as 
\begin{equation}
\mathbf{A}(\boldsymbol{\theta}, \nu) =
\left(\frac{\sin\left(\pi\ \!\frac{d\nu}{c}\ \!\left(\delta - \delta_0\right)\ \!\cos\ \!\delta_0\right)}{\pi\ \!\frac{d\nu}{c}\ \!\left(\delta
  - \delta_0\right)\ \!\cos\ \!\delta_0}\right)^2
\left(\frac{\sin\left(\left(\pi\ \!\frac{b\nu}{c}\right)\ \!\left(\alpha -
  \alpha_0\right)\right)}{\pi\ \!\frac{b\nu}{c}\ \!\left(\alpha - \alpha_0\right)}\right)^2
\end{equation}
% Figure~\ref{fig:antenna-geometry} shows the aperture
%arrangement for OWFA. 
%\begin{figure}
%\begin{center}
%\includegraphics[scale=0.47]{antgeo.eps}
%\end{center}
%\caption{The aperture arrangement for OWFA.}
%\label{fig:antenna-geometry}
%\end{figure}
%The peak of the first sidelobe is at 5\% and 0.5\% at the fourth
%sidelobe.
The $\cos\ \!\delta_0$ factor in the primary beam function arises from
the fact that the aperture is foreshortened in the $d$ direction as
seen from the source at $\delta_0$. This effective reduction in the
aperture size results in a broader primary beam as the declination
increases, as well as reduced sensitivity. In \textbf{Prowess}, by default for
Phase-I, the beam is computed out upto $\sim 18^{\circ}$ from the phase centre in each
direction. This corresponds to three sidelobes north-south, and 10
sidelobes east-west at $\delta_0 = 0^\circ$. The beam is computed and stored as
an array, with a pixel resolution $\sim 1.0' \times
1.0'$, and $2048 \times 2048$ pixels across. The resolution of these maps is
finer than OWFA's resolution of $2^\circ \times 0.1^\circ$.
The simulated foreground maps, elaborated in \citet{Marthi2017a}, are also
computed and stored in an identical sized array. The sinc-squared beam used
here is considered only as a
worst-case scenario, i.e., as having the most pronounced sidelobes. In practice, the beam is a
Gaussian in the east-west dimension as confirmed independently from
slew-scan measurements. To quantify the instrumental chromatic effects  described in Section~\ref{sec:PS_estim}, the sinc-squared beam
would give rise to the strongest signatures. Besides, the effects of representing the primary beam as a
sinc-function in hour-angle are expected to be sub-dominant to the extent that
ORT cannot resolve in that direction: unlike in the declination direction, the
hour-angle term is absent from the Fourier exponent term of the van
Cittert-Zernike theorem.
The full extent of the simulated
primary beam power pattern is shown in Figure~\ref{fig:PRBEAM-vs-dec}
at four different declinations. Having said that, \textbf{Prowess}
can accommodate any definition for the primary beam power pattern, and
need not be constrained to the two-dimensional $\mathrm{sinc}^2$ alone.

\subsection{Sky and model visibilities}\label{ssec:sky-and-model}
The model visibilities are computed as described below. In the model for the
foregrounds, only two components are considered here as they are the most
dominant at 326.5 MHz. The diffuse Galactic
synchrotron foreground dominates the emission from within the
Galaxy at very large scales($ \> 2^{\circ}$), while the extragalactic
sources dominate at scales typically
smaller than a degree\citep{Ali2014}. 
Random realisations of the foreground emission are obtained from the assumed power
spectrum of the emission: the details of how they are generated are
explained in \citet{Marthi2017a}, however a very brief description follows.

For the Galactic diffuse foreground, a random realisation of the angular power
spectrum is generated on a Fourier grid, the values of which are given by 
\begin{equation}
\Delta \tilde{S}(u,\nu)=\left(\frac{\partial B}{\partial T}\right) \sqrt{\frac{\Omega \, C_{\ell}(\nu)}{2}}\left(x+iy\right)
\label{eqn:ran222}
\end{equation}
where $\Delta \tilde{S}(u,\nu)$  represents the Fourier transform of the
intensity fluctuations. $\Omega$ is the solid angle of the sky to be simulated,
$C_\ell$ is the angular power spectrum, $x$ and $y$ are Gaussian random variables
with zero mean and unit variance. An inverse Fourier transform of the populated Fourier
grid now gives a single random realisation of the diffuse foreground.

The extragalactic foreground is similarly obtained from an angular power
spectrum, but it is more involved than the simple Fourier inversion in the
diffuse foreground case. We therefore follow the method of \citet{Gonzalez2005}
by modifying the Poisson density contrast field by the clustering power spectrum
and populating the modified density field with sources drawn from the radio source
counts at 325 MHz \citep{Wieringa1991}.

Once the maps
are available, their individual components are superimposed in
sky-coordinates to obtain the total emission from the sky.
\begin{figure}
\begin{center}
\includegraphics[scale=0.22]{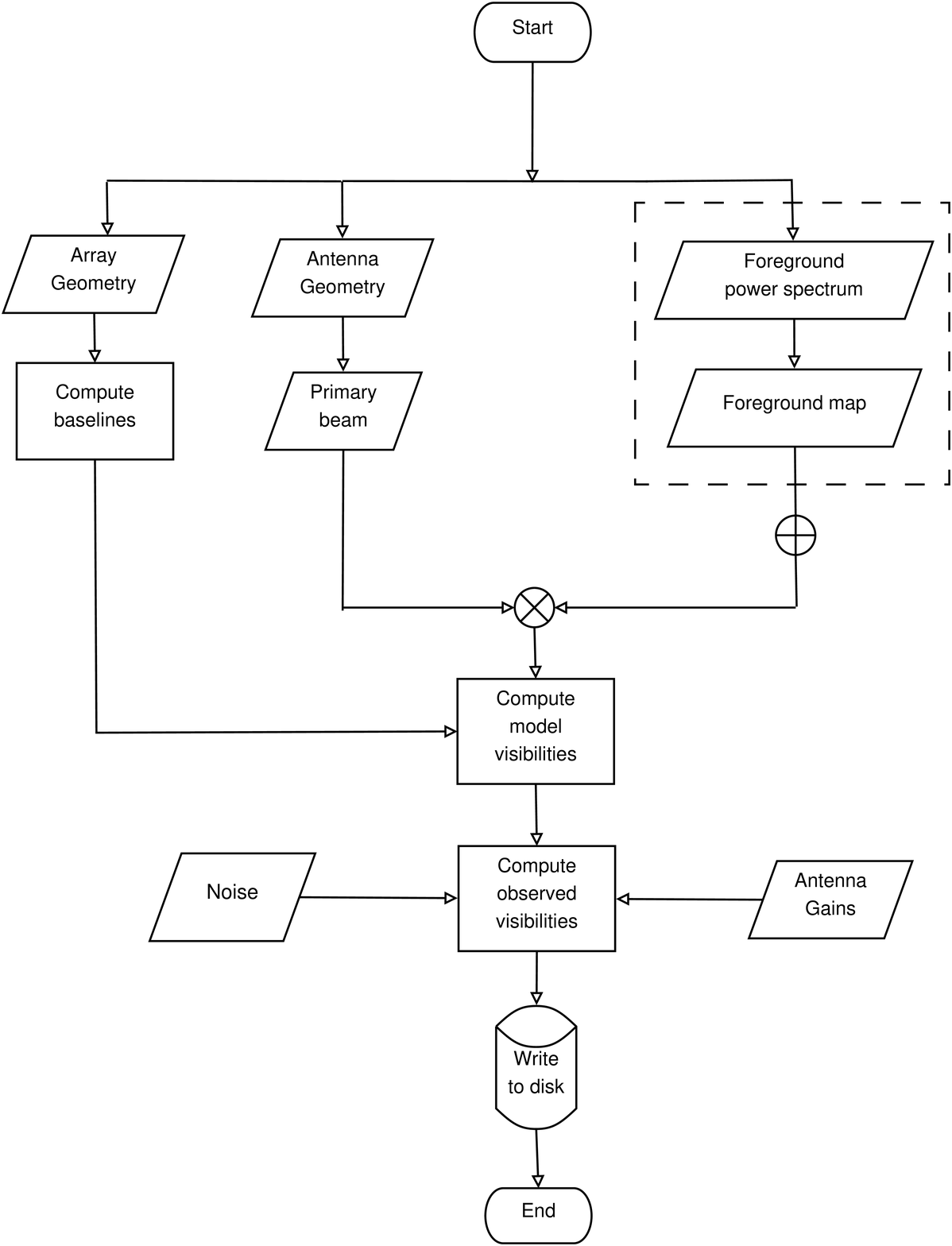}
\end{center}
\caption{This flowchart gives an overall picture of the part of the
  simulator that produces the model and observed visibilities.}
\label{fig:flowchart}
\end{figure}
The specific intensity function is then given by
\begin{equation}
I(\alpha, \delta, \nu) = \Delta I_D(\alpha, \delta, \nu)\ +\ \sum\limits^L_{k=1}
I_k(\alpha_k, \delta_k, \nu)
\end{equation}
where $\Delta I_D(\alpha, \delta, \nu)$ is the fluctuation in the 
specific intensity of the diffuse foreground
emission, and the $L$ distinct extragalactic radio sources are
identified by their co-ordinates $(\alpha, \delta)$ and specific intensities $I_k$. 
The discrete sources are each confined to a pixel, so that the specific intensity
of each source is equal to its flux density. For the diffuse foregrounds, the
simulated maps are already pixelised, and the specific intensity in each pixel
has already been scaled by the solid angle of the pixel. The flux density of the
sky is
\begin{equation}
S(\alpha, \delta, \nu) = \Delta S_D(\alpha, \delta, \nu) +\ \sum\limits^L_{k=1}
S_k(\alpha_k, \delta_k, \nu)
\end{equation}
Since the model visibilities $M(\mathbf{U}, \nu)$ for the
non-redundant set of baselines are obtained as a pixel-by-pixel
Fourier sum of the primary-beam weighted specific intensity
distribution, we finally have
\begin{equation}
\mathbf{M}(\mathbf{U}, \nu) = \sum\limits_{\alpha, \delta}\ S(\alpha, \delta,\nu)\ \mathbf{A}(\alpha, \delta,\nu)\ e^{-i 2\pi \mathbf{U}.\mathbf{\hat{n}}}
\label{eqn:modelvis}
\end{equation}
which is the discretized version of equation~\ref{eqn:model_RIME}.

The observing band is centered at 326.5 MHz with a bandwidth of $\sim39$ MHz split
into 312 channels in the simulations. The frequency resolution in this case is
125 kHz per channel. 
%This resolution is quite well justified in simulations and
%in the final processing of the visibility data. 
Based on our understanding
of the distribution of the neutral gas around the redshift of $z\sim3.35$, it is
expected that the \HI~signal at two redshifted frequencies, separated by more
than $\Delta\nu$ of 1 MHz, decorrelates rapidly\citep[see e.g.][]{
Bharadwaj2005}. This means that with a channel
resolution of 125 kHz, the \HI~signal correlation is adequately sampled over the
1-MHz correlation interval. In reality, the channel resolution is
likely to be much finer($\sim 50 \mathrm{kHz}$), with about 800 channels across the
$\sim$39-MHz band. This is useful for the identification and excision of narrow line
radio frequency interference. Beyond the need to handle RFI, there is no real
incentive to retain the visibility data at this resolution at the cost of
downstream computing and storage requirements. Eventually, we may smooth the
data to a resolution of 125 kHz, keeping in mind the decorrelation bandwidth of
the \HI~signal. The emulator itself is indeed capable of running at any
frequency resolution, including the actual final configuration of the Phase-I and Phase-II
systems. But the 125-kHz resolution used in the simulations allows for rapid
processing, especially if they are to be run repetitively for a wide range of different parameters.

The model visibilities $\mathbf{M(U)}$ need be obtained only
for the set of distinct non-redundant baselines, denoted by
$\mathbf{U}_{|i-j|}$. The observed visibility $V_{ij}$ for a
baseline with antennas $i$ and $j$ depends on the model
visibility for that particular spacing $M(_{|i-j|})$, as well the gains of the individual antennas:
\begin{equation}
V_{ij}\left(U_{|i-j|}\right) = g_i\ g_j^*\ M\left(U_{|i-j|}\right) + N_{ij}\left(\nu\right)
\label{eqn:RIME}
\end{equation}
where $M$ is the model visibility including the primary beam as
described above, $g_i$ and $g_j$ are the complex antenna
gains and $N_{ij}$ is the complex Gaussian random noise equivalent to the system
temperature $T_{\mathrm{sys}}$.  The real and
imaginary parts of the noise $N_{ij}$ in equation~\ref{eqn:RIME} have a
RMS fluctuation
\begin{equation}
\sigma_{ij} = \frac{\sqrt{2}k_B T_{\mathrm{sys}}}{\eta A
  \sqrt{\Delta\nu\Delta t }}
\label{eqn:T_sys}
\end{equation}
per channel, where $k_B$ is the Boltzmann constant, $\eta$ is the
aperture efficiency, $A = b \times d$ is the  aperture area,
$\Delta\nu$ the channel width and $\Delta t$ the integration time. 
The foreground maps give the flux in Jy units at every
pixel, therefore the Fourier sum directly produces the model
visibilities in Jy units as well. The flowchart in Figure~\ref{fig:flowchart} gives a bird's eye
view of the part of the simulator pipeline used to obtain the
visibilities, and summarises the emulator part of \textbf{Prowess}.
The dashed box in the flowchart represents the functionality that simulates the
foreground maps. The emulator also has the functionality to accept an external
FITS image(e.g.~via observations from some other telescope) of the foreground.

\subsection{Redundancy calibration}\label{ssec:calibration}
Due to the highly redundant configuration of OWFA, the number of independent
Fourier modes measured on the sky is small, given by a small number of
baselines, $N_A-1$. All other baselines give copies of these measurements. As a
result, we obtain a system of equations in which the number of unknowns is much
smaller than the number of measurements. This allows us to not only solve for
the gains of the antennas but, unlike routine interferometric calibration, also
solve for the true sky visibilities. This class of calibration algorithms is
called redundancy calibration\citep[see][for examples]{Wieringa1992,
  Liu2010}. \textbf{Prowess} implements a highly efficient, fast and
statistically optimal non-linear least squares minimisation based steepest
descent calibration, that is capable of running in real time. This calibration
algorithm, detailed in \citet{Marthi2014}, is the algorithm of choice for
OWFA. Denoting the antenna gains as $g_i$, the true visibilities as $M_{|i-j|}$
and the observed visibilities as $V_{ij}$, the gains and the true visibilities
are solved for iteratively using the equations
\begin{equation}
g_k^{n+1} = (1 - \alpha) g_k^n + \alpha \mathbf{Q_k}^n
\label{giter}
\end{equation}
and
\begin{equation}
M_{|k-j|}^{n+1} = (1 - \alpha) M_{|k-j|}^n + \alpha \mathbf{R_{kj}}^n
\label{miter}
\end{equation}
where 
\begin{equation}
\mathbf{Q_k} = \frac{\sum\limits_{j \neq
    k} w_{kj} g_j M_{|k-j|}^* V_{kj}}{\sum\limits_{j \neq
    k} w_{kj} |g_j|^2\ |M_{|k-j|}|^2},
\label{itergainsol}
\end{equation}
\begin{equation}
\mathbf{R_{kj}} = \frac{\sum\limits_{j > k} g_k^* g_j V_{kj}}{\sum\limits_{j >
    k}w_{kj} |g_k|^2 |g_j|^2}
\label{itervissol}
\end{equation}
and $0<\alpha<1$ is the step size. The gains and true visibilities at the
present instant are obtained as updates to those obtained in the previous instant.
The simulated visibilities are therefore calibrated using this algorithm
to obtain the true visibilities. Redundancy calibration is a model-independent
calibration procedure; hence the sky model is a natural product of
calibration.

\subsection{Power spectrum estimation}\label{sec:PS_estim}
The calibrated visibility data are processed directly to obtain the power
spectrum. A visibility-based angular power spectrum estimator has been
implemented in \textbf{Prowess}. The estimator has been studied in earlier
works\citep{Begum2006, Datta2007, Choudhuri2014}, but it has been recast for
OWFA.  \textbf{Prowess} implements the estimator to take advantage
of the redundant baseline configuration of OWFA, allowing us to add the
visibilities from all the redundant copies of a baseline of a given length. The
implementation of the estimator is detailed in \citet{Marthi2017a}. 
Denoting the calibrated visibility as $\mathcal{V}(\mathbf{U}_n,\nu)$,
let us define the quantity 
\begin{equation}
\mathcal{V}^\prime(\mathbf{U}_n,\nu) = \sum_{i=0}^{N_n} |\mathcal{V}^{(i)}(\mathbf{U}_{n}, \nu)|^2
\end{equation}
We can now define our estimator, $\mathbf{S}_2$, as
\begin{equation}
\mathbf{S}_2(\mathbf{U}_n, \nu_i, \nu_j) = \frac{\mathcal{V}(\mathbf{U}_n,
  \nu_i)\mathcal{V}^*(\mathbf{U}_n, \nu_j) - \delta_{ij}\mathcal{V}^\prime(\mathbf{U}_n,
  \nu_i)}{N_n^2 - \delta_{ij}N_n}
\label{eqn:S2_compute}
\end{equation}
The second term in the numerator corrects for the bias contributed by the
self-correlated noise.
For an N-channel visibility dataset, the real-valued $\mathbf{S}_2(\mathbf{U}_n,
\nu_i, \nu_j)$ is a $N \times N$ matrix as implemented in \textbf{Prowess}.
This representation in the $\nu-\nu$ plane has its
advantages in the study of systematics as it retains the full spectral
information of the foregrounds and those introduced by the instrument. The
estimator matrix cube is available in FITS format that can be visulaised either
through standard FITS visualisation programs or through the tool implemented in
\textbf{Prowess}.

The study of foreground power spectra through simulations, enabled by
\textbf{Prowess}, inform us that per-baseline spectral features arising in the estimator within the
observing band are caused by (i) the effect of the
frequency-dependent primary beam and (ii) the frequency dependence of the
baseline vector. These effects have been studied in some detail in the context
of EoR experiments, such as by \citet{Vedantham2012} for example. We have
carried out exhaustive studies of instrumental chromatic effects on the power
spectrum for OWFA \citep{Marthi2017a}. Importantly, simulations using
\textbf{Prowess} appear to indicate that the instrinsic chromatic properties of
the sky are less important, but such chromatic effects are compounded by the
chromatic instrument response. These effects would eventually play a limiting
role in the detectability of the \HI~signal.

\subsection{Data visualisation}
The simulated visibilities are written into a FITS file in the UVFITS
data format. This is a standard format for reading and writing the
radio interferometric visibility data, and is the format being used at the GMRT. However, \textbf{Prowess} has
its own interface that helps in visualising the visibility data, which is
explained here. The data are available in time-baseline-frequency
order. That is, the coarsest data identifier is the record number,
which is tagged to the timestamp. At each timestamp, all the $^{N_A}C_2$
baselines are sorted in a specific order, indexed by the FITSbl
\begin{figure}
\begin{center}
\includegraphics[scale=0.55]{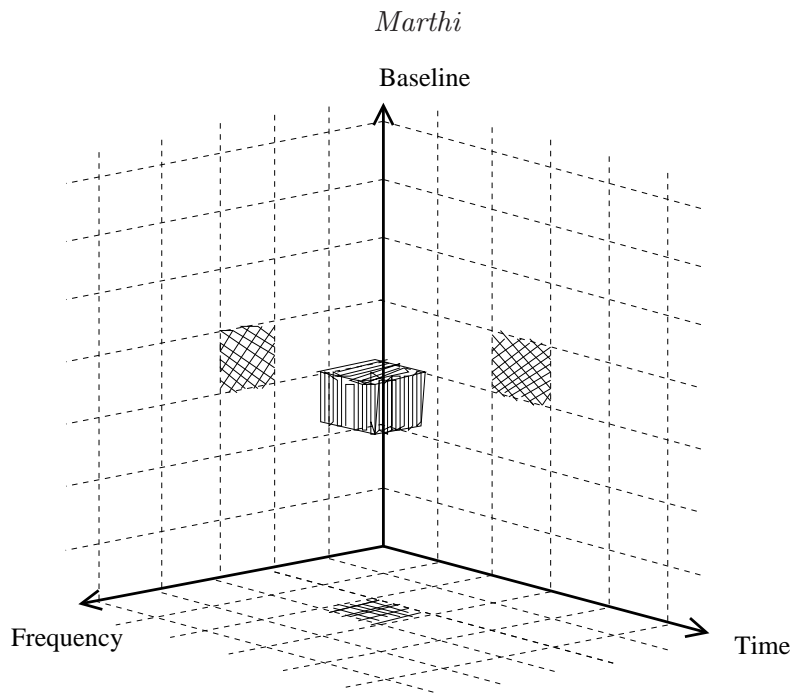}
\end{center}
\caption{The display buffer with gridded Time, Frequency and Baseline axes,
  where the complex visibility resides.}
\label{fig:displaybuf}
\end{figure}
number shown in Figure~\ref{fig:baselines.info}. Each baseline has $N$
channels, and each channel has a real and imaginary number for the
visibility, and an associated weight. The visibility data therefore reside in a
\begin{figure}
\begin{center}
\includegraphics[scale=0.27,angle=270]{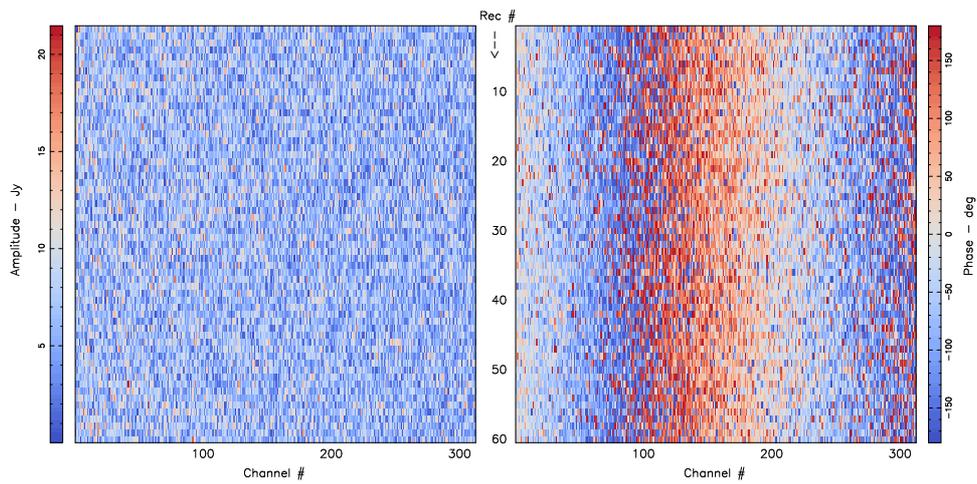}
\end{center}
\caption{A time-frequency view of the visibility data on the $T-N$
  plane for $B=1$.}
\label{fig:imagtifr}
\end{figure}
\begin{figure}
\begin{center}
\includegraphics[scale=0.27,angle=270]{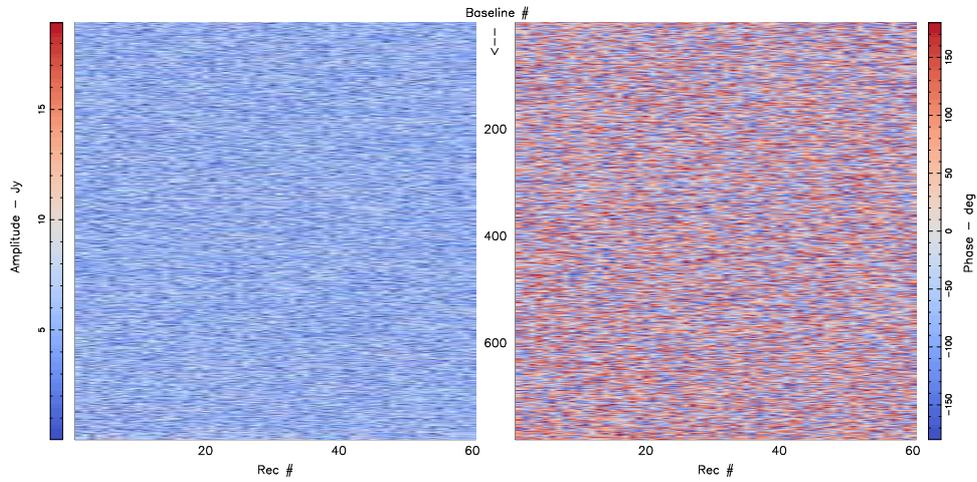}
\end{center}
\caption{A time-baseline view of the visibility data on the $T-B$
  plane for $N=156$.}
\label{fig:imagtibl}
\end{figure}
\begin{figure}
\begin{center}
\includegraphics[scale=0.27,angle=270]{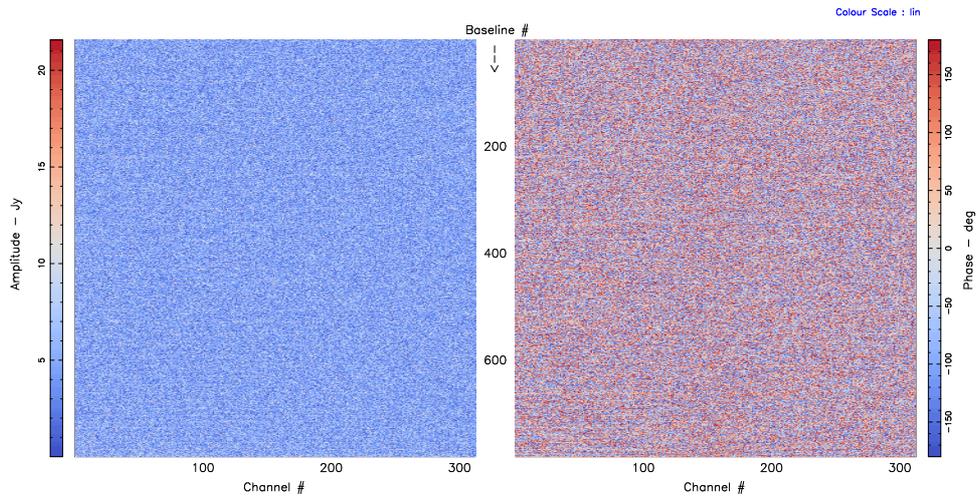}
\end{center}
\caption{A frequency-baseline view of the visibility data on the $N-B$
  plane for $T=30$.}
\label{fig:imagfrbl}
\end{figure}
gridded three-dimensional co-ordinate system where the three axes are time,
baseline and frequency. There are $T \times B \times N$ grid cells,
where $T$ is the number of records, $B$ is the number of baselines and
$N$ is the number of channels, shown in Figure~\ref{fig:displaybuf}. This
arrangement is very convenient; therefore
the native structure that holds these visibility data in a display
buffer is similarly defined. For visualising the data, they are therefore
read from the FITS record into the display buffer. The visualisation
programs hence access one of the planes parallel to the $T-B$, $B-N$
or $T-N$ planes. Figures~\ref{fig:imagtifr}, ~\ref{fig:imagtibl} and
~\ref{fig:imagfrbl} show the simulated visibility accessed
from the display buffer for an example run with a particular
realisation for the diffuse Galactic foreground, with the 312-channel, 40-antenna,
780-baseline Phase-I observed for 60 seconds with a record being
written to disk every second. There is an option to dynamically switch
between a real/imaginary or an amplitude/phase view, and dynamically switch
between linear, square-root and logarithmic image transfer functions. A dynamic
zooming feature is available in these interfaces as well. Besides the
colour-coded plan view, these plane data from the display buffer can
also be viewed in a line-plot format, again, with the
option to dynamically switch between real/imaginary and
amplitude/phase view formats.
%\newpage
\section{The observatory data processing pipeline}
\textbf{Prowess} is not an emulator alone, as the name may suggest. It
was described in Section~\ref{sec:rationale} why the emulator was
conceived: it serves the dual purpose of an emulator and
would potentially become the standard post-correlation data processing
pipeline at the observatory. It would be useful now to dwell a little
on the data processing pipeline aspect of \textbf{Prowess} and state
what functionality it is meant to provide.

The data pooled from the antennas would terminate in the eight
high-performance compute nodes through eleven copper ethernet links
each. These nodes correlate the signal from every pair of antennas and
accumulate the products upto an interval of time, usually
programmable. The typical integration time, called the Long Term
Accumulation(LTA), is of the order of $1-10$ seconds. 
%However, for
%ease of sanity checks and other logistical considerations, this LTA
%visibility data is usually written to the disks in a native
%format. THe details of the native format are not important nor
%relevant to the discussion presented here. It is possible to access
%these data either from the disk, which is more expensive, or read them
%out directly from the (shared)memory, which is quicker. Therefore, we
%envisage two different interfaces - one to read from shared memory and
%another from the disk - but write the data to disk in a single common
%format. This format could, for example, be the standard FITS
%format. I have already argued in Section~\ref{ssec:architecture} why
%the FITS format is advantageous and more attractive than the MS
%format. 
Once the data are available in FITS format, \textbf{Prowess} can
completely take over downstream processing, which include
calibration and power spectrum estimation. The uncalibrated FITS
data can as well be stored in the disks for offline calibration. The enormous
redundancy of the measurements and the structure of
OWFA are best exploited by calibrating the visibilities using the non-linear least
squares redundancy calibration algorithm, described in
Section~\ref{ssec:calibration}. 
The calibrated visibilities are later processed to obtain the power spectrum of observed
sky.  

\section{Summary}
A software model that captures the instrumental and geometric
details of OWFA has been described. This detailed model is an important aid to understanding the
systematics introduced by the instrument and to make robust and meaningful
predictions for the foregrounds and the \HI~signal. The programming philosophy
allows for modular function definitions and easy addition of new
functionality. The suite features a rich and interactive visual environment to
play back the visibility data. These programs comprise not just an emulator for OWFA, but they
are also designed to serve as observatory data analysis software. Prowess has greatly
aided our understanding of the instrument and the systematics expected in the OWFA cosmology
experiment.

\bibliographystyle{mnras}
\bibliography{mylist.bib}

\begin{thebibliography}{30}
\expandafter\ifx\csname natexlab\endcsname\relax\def\natexlab#1{#1}\fi

\bibitem[{{Ali} \& {Bharadwaj}(2014)}]{Ali2014}
{Ali} S.~S., {Bharadwaj} S., 2014, \japa, 35, 157

\bibitem[{{Ali}, {Bharadwaj} \& {Chengalur}(2008){Ali}, {Bharadwaj}, \&
  {Chengalur}}]{Ali2008}
{Ali} S.~S., {Bharadwaj} S., {Chengalur} J.~N., 2008, \mnras, 385, 2166

\bibitem[{{Bandura} {et~al}\mbox{.}(2014){Bandura}, {Addison}, {Amiri}, {Bond},
  {Campbell-Wilson}, {Connor}, {Cliche}, {Davis}, {Deng}, {Denman}, {Dobbs},
  {Fandino}, {Gibbs}, {Gilbert}, {Halpern}, {Hanna}, {Hincks}, {Hinshaw},
  {H{\"o}fer}, {Klages}, {Landecker}, {Masui}, {Mena Parra}, {Newburgh}, {Pen},
  {Peterson}, {Recnik}, {Shaw}, {Sigurdson}, {Sitwell}, {Smecher}, {Smegal},
  {Vanderlinde}, \& {Wiebe}}]{Bandura2014}
{Bandura} K. {et~al.}, 2014, in \procspie, Vol. 9145, Ground-based and Airborne
  Telescopes V, p. 914522

\bibitem[{{Begum}, {Chengalur} \& {Bhardwaj}(2006){Begum}, {Chengalur}, \&
  {Bhardwaj}}]{Begum2006}
{Begum} A., {Chengalur} J.~N., {Bhardwaj} S., 2006, \mnras, 372, L33

\bibitem[{{Bharadwaj} \& {Ali}(2005)}]{Bharadwaj2005}
{Bharadwaj} S., {Ali} S.~S., 2005, \mnras, 356, 1519

\bibitem[{{Bharadwaj}, {Sarkar} \& {Ali}(2015){Bharadwaj}, {Sarkar}, \&
  {Ali}}]{Bharadwaj2015}
{Bharadwaj} S., {Sarkar} A.~K., {Ali} S.~S., 2015, \japa, 36, 385

\bibitem[{{Chen}(2011)}]{Chen2011}
{Chen} X., 2011, Scientia Sinica Physica, Mechanica Astronomica, 41,
  1358

\bibitem[{{Choudhuri} {et~al}\mbox{.}(2014){Choudhuri}, {Bharadwaj}, {Ghosh},
  \& {Ali}}]{Choudhuri2014}
{Choudhuri} S., {Bharadwaj} S., {Ghosh} A., {Ali} S.~S., 2014, \mnras, 445,
  4351

\bibitem[{{Datta}, {Choudhury} \& {Bharadwaj}(2007){Datta}, {Choudhury}, \&
  {Bharadwaj}}]{Datta2007}
{Datta} K.~K., {Choudhury} T.~R., {Bharadwaj} S., 2007, \mnras, 378, 119

\bibitem[{{Gonz{\'a}lez-Nuevo}, {Toffolatti} \&
  {Arg{\"u}eso}(2005){Gonz{\'a}lez-Nuevo}, {Toffolatti}, \&
  {Arg{\"u}eso}}]{Gonzalez2005}
{Gonz{\'a}lez-Nuevo} J., {Toffolatti} L., {Arg{\"u}eso} F., 2005, \apj, 621, 1

\bibitem[{{Iacobelli} {et~al}\mbox{.}(2014){Iacobelli}, {Burkhart},
  {Haverkorn}, {Lazarian}, {Carretti}, {Staveley-Smith}, {Gaensler},
  {Bernardi}, {Kesteven}, \& {Poppi}}]{Iacobelli2014}
{Iacobelli} M. {et~al.}, 2014, \aap, 566, A5

\bibitem[{{Iacobelli}, {Haverkorn} \& {Katgert}(2013){Iacobelli}, {Haverkorn},
  \& {Katgert}}]{Iacobelli2013a}
{Iacobelli} M., {Haverkorn} M., {Katgert} P., 2013, \aap, 549, A56

\bibitem[{{Iacobelli} {et~al}\mbox{.}(2013){Iacobelli}, {Haverkorn},
  {Orr{\'u}}, {Pizzo}, {Anderson}, {Beck}, {Bell}, {Bonafede}, {Chyzy},
  {Dettmar}, {En{\ss}lin}, {Heald}, {Horellou}, {Horneffer}, {Jurusik},
  {Junklewitz}, {Kuniyoshi}, {Mulcahy}, {Paladino}, {Reich}, {Scaife}, {Sobey},
  {Sotomayor-Beltran}, {Alexov}, {Asgekar}, {Avruch}, {Bell}, {van Bemmel},
  {Bentum}, {Bernardi}, {Best}, {B{\i}rzan}, {Breitling}, {Broderick}, {Brouw},
  {Br{\"u}ggen}, {Butcher}, {Ciardi}, {Conway}, {de Gasperin}, {de Geus},
  {Duscha}, {Eisl{\"o}ffel}, {Engels}, {Falcke}, {Fallows}, {Ferrari},
  {Frieswijk}, {Garrett}, {Grie{\ss}meier}, {Gunst}, {Hamaker}, {Hassall},
  {Hessels}, {Hoeft}, {H{\"o}randel}, {Jelic}, {Karastergiou}, {Kondratiev},
  {Koopmans}, {Kramer}, {Kuper}, {van Leeuwen}, {Macario}, {Mann}, {McKean},
  {Munk}, {Pandey-Pommier}, {Polatidis}, {R{\"o}ttgering}, {Schwarz}, {Sluman},
  {Smirnov}, {Stappers}, {Steinmetz}, {Tagger}, {Tang}, {Tasse}, {Toribio},
  {Vermeulen}, {Vocks}, {Vogt}, {van Weeren}, {Wise}, {Wucknitz}, {Yatawatta},
  {Zarka}, \& {Zensus}}]{Iacobelli2013b}
{Iacobelli} M. {et~al.}, 2013, \aap, 558, A72

\bibitem[{{Kemball} \& {Wieringa}(2000)}]{MS}
{Kemball} A.~J., {Wieringa} M.~H., 2000, https://casacore.github.io/casacore-notes/229.html

\bibitem[{{Liu} {et~al}\mbox{.}(2010){Liu}, {Tegmark}, {Morrison},
  {Lutomirski}, \& {Zaldarriaga}}]{Liu2010}
{Liu} A., {Tegmark} M., {Morrison} S., {Lutomirski} A., {Zaldarriaga} M., 2010,
  \mnras, 408, 1029

\bibitem[{{Marthi} {et~al}\mbox{.}(2017){Marthi}, {Chatterjee}, {Chengalur}, \&
  {Bharadwaj}}]{Marthi2017a}
{Marthi} V.~R., {Chatterjee} S., {Chengalur} J., {Bharadwaj} S., 2017, \mnras,
under review

\bibitem[{{Marthi} \& {Chengalur}(2014)}]{Marthi2014}
{Marthi} V.~R., {Chengalur} J., 2014, \mnras, 437, 524

\bibitem[{{Pober} {et~al}\mbox{.}(2013){Pober}, {Parsons}, {DeBoer},
  {McDonald}, {McQuinn}, {Aguirre}, {Ali}, {Bradley}, {Chang}, \&
  {Morales}}]{Pober2013b}
{Pober} J.~C. {et~al.}, 2013, \aj, 145, 65

\bibitem[{{Prasad} \& {Subrahmanya}(2011)}]{Prasad2011}
{Prasad} P., {Subrahmanya} C.~R., 2011, \exa, 31, 1

\bibitem[{{Price}, {Barsdell} \& {Greenhill}(2015){Price}, {Barsdell}, \&
  {Greenhill}}]{Price2015}
{Price} D.~C., {Barsdell} B.~R., {Greenhill} L.~J., 2015, in Astronomical
  Society of the Pacific Conference Series, Vol. 495, Astronomical Data
  Analysis Software an Systems XXIV (ADASS XXIV), {Taylor} A.~R., {Rosolowsky}
  E., eds., p. 531

\bibitem[{{Santos}, {Cooray} \& {Knox}(2005){Santos}, {Cooray}, \&
  {Knox}}]{Santos2005}
{Santos} M.~G., {Cooray} A., {Knox} L., 2005, \apj, 625, 575

\bibitem[{{Sarkar}, {Bharadwaj} \& {Ali}(2017){Sarkar}, {Bharadwaj}, \&
  {Ali}}]{Sarkar2017a}
{Sarkar} A.~K., {Bharadwaj} S., {Ali} S.~S., 2017, \japa, special issue on OWFA

\bibitem[{{Subrahmanya}, {Manoharan} \& {Chengalur}(2017){Subrahmanya},
  {Manoharan}, \& {Chengalur}}]{Subrahmanya2017a}
{Subrahmanya} C.~R., {Manoharan} P.~K., {Chengalur} J.~N., 2017, \japa, special
issue on OWFA

\bibitem[{{Subrahmanya} {et~al}\mbox{.}(2017){Subrahmanya}, {Prasad}, {Girish},
  {Somasekhar}, {Manoharan}, \& {Amit Mittal}}]{Subrahmanya2017b}
{Subrahmanya} C.~R., {Prasad} P., {Girish} B.~S., {Somasekhar} R., {Manoharan}
  P.~K., {Amit Mittal} S.~G., 2017, \japa, special issue on OWFA

\bibitem[{{Swarup} {et~al}\mbox{.}(1971){Swarup}, {Sarma}, {Joshi}, {Kapahi},
  {Bagri}, {Damle}, {Ananthakrishnan}, {Balasubramanian}, {Bhave}, \&
  {Sinha}}]{Swarup1971}
{Swarup} G. {et~al.}, 1971, Nature Physical Science, 230, 185

\bibitem[{{Vedantham}, {Udaya Shankar} \& {Subrahmanyan}(2012){Vedantham},
  {Udaya Shankar}, \& {Subrahmanyan}}]{Vedantham2012}
{Vedantham} H., {Udaya Shankar} N., {Subrahmanyan} R., 2012, \apj, 745, 176

\bibitem[{{Wells}, {Greisen} \& {Harten}(1981){Wells}, {Greisen}, \&
  {Harten}}]{FITS}
{Wells} D.~C., {Greisen} E.~W., {Harten} R.~H., 1981, \aaps, 44, 363

\bibitem[{{Wieringa}(1991)}]{Wieringa1991}
{Wieringa} M.~H., 1991, PhD thesis, Rijksuniversiteit Leiden, (1991)

\bibitem[{{Wieringa}(1992)}]{Wieringa1992}
{Wieringa} M.~H., 1992, Experimental Astronomy, 2, 203

\bibitem[{{Xu}, {Wang} \& {Chen}(2015){Xu}, {Wang}, \& {Chen}}]{Xu2015}
{Xu} Y., {Wang} X., {Chen} X., 2015, \apj, 798, 40

\end{thebibliography}
\label{lastpage}

\end{document}